\documentclass[12pt,amstex,aps,amsfonts,prsty,preprint]{article} 
\usepackage[dvips]{graphicx}
\usepackage{amssymb}
\usepackage{amsmath}
\usepackage{a4wide}
\usepackage{citesort}
\newcommand{\be}{\begin{equation}}
\newcommand{\ee}{\end{equation}}

\begin{document}

\begin{titlepage}

\begin{center}

{\Large \bf Single-Species Reactions
on
a Random Catalytic Chain.}

\vspace{0.1in}

{\Large  G.Oshanin$^{1}$ and S.F.Burlatsky$^{2}$}

\vspace{0.1in}

{$^{1}$ \sl Laboratoire de Physique Th{\'e}orique des Liquides, \\
Universit{\'e} Paris 6, 4 Place Jussieu, 75252 Paris, France}

\vspace{0.1in}

{$^{2}$ \sl United Technologies Research Center,\\
United Technologies Corporation,\\ 
411 Silver Lane, 129-21 East Hartford, CT 06108, USA
}

\vspace{0.1in}

\begin{abstract}
We present an exact solution for 
a catalytically-activated  
annihilation $A + A \to 0$  
reaction taking place on a one-dimensional 
chain 
in which some segments (placed at 
random, with mean concentration $p$) possess 
special, catalytic properties. 
Annihilation 
reaction takes place, as soon as
any two $A$ particles 
land from the reservoir
 onto two vacant sites at the
extremities of the 
 catalytic segment, or when any 
$A$ particle 
lands onto a vacant site on 
a catalytic segment
while the site at the 
other extremity of this
segment is already occupied 
by another $A$ particle.   
We find that
the disorder-average pressure $P^{(quen)}$ per site of 
such a chain 
is given by
$P^{(quen)} =  P^{(lan)} + 
\beta^{-1} F$, where
 $P^{(lan)} = \beta^{-1} \ln(1+z)$ is
the Langmuir adsorption 
pressure, ($z$ being the activity and
$\beta^{-1}$ - the temperature),  while 
$\beta^{-1} F$ is the 
reaction-induced contribution, which 
can be expressed,
under appropriate change of notations,  
as the  Lyapunov exponent
for the product of $2 \times 2$ random matrices, 
obtained exactly by Derrida and Hilhorst (J. Phys. A {\bf 16}, 2641 (1983)).
Explicit asymptotic 
formulae for the particle
mean density and the compressibility are also presented. 
\end{abstract}

\vspace{0.1in}
\end{center}

\end{titlepage}

\section{Introduction.}

Catalytically-activated reactions (CARs), i.e. 
reactions between 
chemically inactive molecules
which recombine only
when some third substance - the catalytic substrate - is present, 
are widespread in nature \cite{1a,7a}. 
Recently, 
 such reactions 
have attracted a considerable attention 
following an early observation  \cite{3a}  of remarkable non-mean-field behavior
exhibited by
a specific
reaction  -  the CO-oxidation in the presence of 
metal surfaces with  catalytic properties 
\cite{1a,7a}. An extensive analysis of this CAR
has substantiated 
the emergence of an 
essentially different behavior as compared to the
predictions of the classical, formal-kinetics scheme
and have shown that under certain conditions
such
collective phenomena 
as phase transitions
or the formation of
bifurcation patterns may take place \cite{3a}.
Prior to these works on catalytic systems,  
anomalous behavior was
amply demonstrated 
in other 
schemes \cite{4a}, involving reactions on contact between two particles 
 at any point of the reaction volume 
(i.e., the "completely" catalytic sysems).
It
was realized \cite{4a} 
that the departure from the text-book, formal-kinetic predictions is
due to 
 many-particle
effects, associated with fluctuations in the spatial
 distribution of the reacting species.  
This suggests that 
similarly to such
 "completely" catalytic
reaction 
schemes,
 the behavior of the 
CARs
 may be influenced
 by many-particle
effects.

Apart from 
the  many-particle effects,  
behavior of the CARs in practically involved systems
might be affected
by the very 
structure of the catalytic substrate, which is often
not
well-defined geometrically, 
but must be viewed as an assembly
of  mobile or localized
 catalytic
sites or islands, whose spatial distribution 
is complex \cite{1a}.
  Metallic catalysts, for instance, 
are often disordered
compact aggregates, the building blocks 
of which are imperfect crystallites
with broken faces, kinks and steps. 
Another example is furnished by porous materials with convoluted surfaces,
such as, e.g.,
silica, alumina or carbons. Here the effective catalytic
 substrate is also only a portion of the total surface area 
because of the selective participation 
of different surface sites to the reaction.
Finally, 
for  liquid-phase 
CARs the 
catalyst can consist of active  groups attached 
to polymer chains in solution.

Such complex morphologies render the theoretical analysis difficult.
As yet, only empirical approaches have been used
to account for the impact of the geometrical
complexity on  the behavior of the CARs, based mostly 
 on  heuristic concepts of
effective reaction order 
or on phenomenological 
generalizations of the formal-kinetic "law of mass action" 
(see, e.g. Refs.\cite{1a} and \cite{7a} for more details). 
In this regard, analytical solutions of even somewhat 
idealized or simplified 
models, such as, for instance, 
the ones proposed
in Refs.\cite{3a}, are already 
highly desirable since such studies may
  provide an understanding 
of the  effects of different factors on 
the properties of the CARs.

\begin{figure}[ht]
\begin{center}
\includegraphics*[scale=0.5]{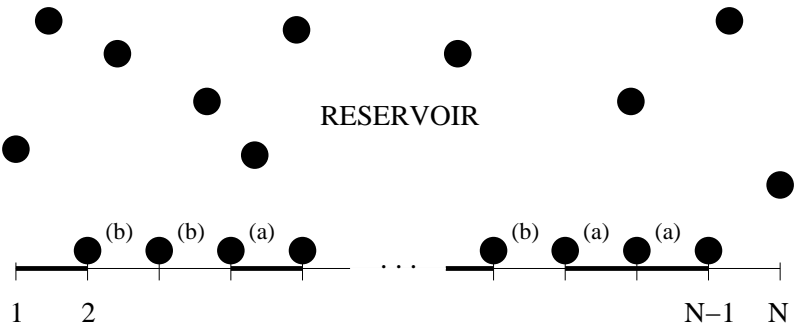}
\caption{\label{Fig1} {\small One-dimensional lattice of 
adsorption sites in contact with a
reservoir. 
Filled circles denote hard-core $A$ particles. 
Thick black lines denote the
segments with catalytic properties. 
(a) denotes a "forbidden" particle
configuration, which corresponds 
to immediate reaction. (b) depicts the situation in which two 
neighboring $A$ particles may harmlessly coexist. 
}}
\end{center}
\end{figure}

In this paper we study 
a catalytically-activated annihilation
$A + A \to 0$ reaction 
in  a simple, one-dimensional
model 
with random distribution 
of the catalyst, appropriate to the 
just mentioned situation
with the catalytically-activated reactions
 on polymer chains. We present here an
 exact 
solution for this model with quenched
random distribution of the catalyst
and show that despite its apparent simplicity
it exhibits an 
interesting non-trivial behavior.  
We note finally 
that  
kinetics of $A + A \to 0$ reactions involving diffusive $A$ particles
which react upon encounters on randomly placed catalytic sites 
has been discussed already in Refs.\cite{bur,oshan1,tox} and \cite{nech}, and a rather
surprising behavior has been found, especially in low-dimensional systems. 
Additionally, 
steady-state properties of $A + A \to 0$ 
reactions between immobile $A$ particles
with long-range reaction probabilities
in systems with external particles input have been 
presented in Refs.\cite{sander} and \cite{deem}
and revealed non-trivial ordering phenomena with 
anomalous input intensity dependence of the 
mean 
particle density, which agrees with 
experimental observations \cite{ben}.

\section{The model.}

Consider a one-dimensional regular lattice of unit spacing 
comprising $N$ adsorption sites. The lattice is 
in contact with a 
reservoir of identical, non-interacting hard-core
$A$ particles (see, Fig.1) - a vapor phase, which  is steadily 
maintained at a constant pressure.

The $A$ particles from
the vapor phase can adsorb onto vacant adsorption sites
and desorb back to the reservoir. 
The occupation of the "i"-th adsorption site is
described by the Boolean variable $n_i$, such that
\begin{equation}
n_i = \left\{\begin{array}{ll}
1,     \mbox{ if the "i"-th site is occupied,} \nonumber\\
0,     \mbox{ otherwise.}
\end{array}
\right.
\end{equation}

Suppose next that some of the segments - intervals between neighboring
adsorption sites possess 
"catalytic" properties (thick black lines 
in Fig.1)
in the
sense that they induce an immediate
 reaction $A + A \to 0$, as soon as
two $A$ particles land onto two vacant sites at the
extremities of the  catalytic segment, or an 
$A$ particle lands onto a vacant site at one extremety of the catalytic segment
while the site at the 
other extremity of this 
segment is already occupied 
by another $A$ particle. Two reacted $A$ particles 
instantaneously leave the lattice (desorb
back to the reservoir). Any two $A$ particle adsorbed at extremities of a 
non-catalytic segment harmlessly coexist.

To specify the positions of the 
catalytic segments, we introduce a Boolean variable 
$\zeta_i$, so that $\zeta_0 = \zeta_N = 0$ and 
\begin{equation}
\zeta_i = \left\{\begin{array}{ll}
1,     \mbox{if the $i$-t interval is catalytic, $i = 1,2, \ldots , N - 1$, } \nonumber\\
0,     \mbox{ otherwise.}
\end{array}
\right.
\end{equation}
In what follows we suppose that ${\zeta_i}$ are independent, identically distributed 
quenched random variables
with distribution
\begin{equation}
\rho(\zeta) = p \delta(\zeta - 1) + (1 - p) \delta(\zeta).
\end{equation} 

Now, for a given distribution of the  catalytic segments, the 
partition function $Z_N(\zeta)$ of 
the system under study can be written as follows: 
\begin{equation}
\label{partition}
Z_N(\zeta) = \sum_{\{n_i\}} 
z^{\sum_{i = 1}^N  n_i} \; \prod_{i = 1}^{N - 1} \Big(1 - \zeta_i \; n_i \; n_{i+1} \Big), 
\end{equation}
where 
the summation $\sum_{\{n_i\}}$ extends over all 
possible configurations
$\{n_i\}$, while 
\begin{equation}
z = \exp(\beta \mu)
\end{equation}
is the activity and $\mu$ - the chemical potential.
Note that $Z_N(\zeta)$ in Eq.(\ref{partition})   
is a functional of the
configuration $\zeta = \{\zeta_i\}$. 

It is worth-while to remark 
that $Z_N(\zeta)$ can also be  thought of as a 
one-dimensional version of
models describing adsorption 
of hard-molecules \cite{0,01,001,0001,1,2,3,heringa},
i.e. adsorption limited by the 
"kinetic" constraint 
that any two molecules
can neither occupy the same 
site nor appear on the neighboring sites. The most celebrated 
examples of such models are furnished  by
the so-called "hard-squares" model \cite{0,01,001,0001,1},  
or by the  "hard-hexagons" model first 
solved exactly by Baxter \cite{3}.
In our case of the CARs on random catalytic substrates
the nearest-neighbor exclusion 
constraint is
introduced only locally, at 
some specified, randomly distributed intervals. 
Such locally frustrated models of 
random reaction/adsorption
thus represent a natural and meaningful 
generalization
of the well-studied exclusion models over systems with disorder. 
Of course, in this context 
two-dimensional situations are of most interest, but
nonetheless 
it might be instructive to 
find examples of such models
which can be solved exactly in one dimension.

Our main goal here is 
to calculate 
the disorder-average pressure
per site:
\begin{equation}
\label{3}
P^{(quen)} = \frac{1}{\beta} 
\lim_{N \to \infty} \frac{1}{N} \Big<\ln(Z_N(\zeta)) \Big>_{\zeta},
\end{equation}
where the angle brackets with the subscript $\zeta$  
denote 
averaging over all possible configurations $\{\zeta_i\}$. 
Once $P^{(quen)}$ is obtained,  
all other pertinent thermodynamic properties can be readily evaluated
by differentiating $P^{(quen)}$ with respect to the chemical potential $\mu$; 
in particular,
the disorder-average mean particle density $n^{(quen)}$ will be given by
\begin{equation}
\label{dens}
n^{(quen)} = \frac{\partial }{\partial \mu} P^{(quen)},
\end{equation}
while the compressibility $k_T$
obeys 
\begin{equation}
k_T^{(quen)} = \frac{1}{\Big(n^{(quen)}\Big)^2} \frac{\partial n^{(quen)}}{\partial \mu}. 
\end{equation}

To close this section, we display the results 
corresponding to two "regular" cases: namely, when $p = 0$ and $p = 1$, 
which will serve us in what follows as
some benchmarks. In the $p = 0$ all sites are decoupled, and 
one has the  Langmuir results:
\begin{equation}
\label{densi}
P^{(lan)} = \frac{1}{\beta} \ln(1 + z), \;\;\; n^{(lan)} = \frac{z}{1 + z}, \;\;\; \text{and} \;\;\;
\beta^{-1} k_T^{(lan)} = \frac{1}{z}
\end{equation}
The "regular" case when $p = 1$ is a bit 
less trivial, but the solution can be still straightforwardly obtained. 
In this case, we have
\begin{equation}
\label{ord}
P^{(reg)} = \frac{1}{\beta} \ln\Big(\frac{\sqrt{1 + 4 z} + 1}{2}\Big), \;\;\; 
n^{(reg)} = 1 - \frac{2 z}{1 + 4 z - \sqrt{1 + 4 z}},
\end{equation}
and
\begin{equation}
\beta^{-1} k_T^{(reg)} = \frac{2 z}{\sqrt{1 + 4 z} (1 + 2 z - \sqrt{1 + 4 z})}.
\end{equation}
Note that in the $p=1$ case (the completely catalytic system) 
the mean particle density tends to $1/2$ as $z \to \infty$ 
(compared to $n^{(lan)} \to 1$ behavior 
observed for the Langmuir case), which means that the 
adsorbent undergoes "ordering" transition
and particles distribution on the lattice becomes 
periodic revealing a spontaneous symmetry breaking between 
two sublattices. 
In the limit $z \to \infty$ 
the compressibility 
vanishes
as $k_T^{(reg)} \propto 1/\sqrt{z}$ compared to the Langmuir behavior 
$k_T^{(lan)} \propto 1/z$.

\section{Recursion relations for $Z_N(\zeta)$.}

Let us first introduce an auxiliary, constrained partition 
function of the form
\begin{equation}
Z_N'(\zeta) = \left. Z_N(\zeta)\right|_{ n_{N}  = 1} = z \sum_{\{n_i\}} 
z^{\sum_{i = 1}^{N-1}  n_i} \; \prod_{i = 1}^{N - 2} 
\Big(1 - \zeta_i \; n_i \; n_{i+1} \Big)  
\Big(1 - \zeta_{N - 1} \; n_{N-1} \Big),
\end{equation}
i.e. $Z_N'(\zeta)$ stands for
 the partition function of a system with  
fixed set $\zeta = \{\zeta_i\}$ and fixed occupation of the site 
$i = N$, $n_N = 1$. 
Evidently, we have that
\begin{equation}
\label{1}
Z_N(\zeta) = Z_{N-1}(\zeta) + Z_N'(\zeta). 
\end{equation}
Next, considering two possible values 
of the occupation variable $n_{N - 1}$, 
i.e. $n_{N - 1} = 0$ and $n_{N - 1} = 1$, we find that $Z_N'(\zeta)$ can be expressed through $Z_ {N-2}(\zeta)$ and $Z_{N-1}'(\zeta)$
as
\begin{eqnarray}
\label{2}
Z_N'(\zeta) &=&  z \sum_{\{n_i\}} 
z^{\sum_{i = 1}^{N-2}  n_i} \; \prod_{i = 1}^{N - 3} \Big(1 - \zeta_i \; n_i \; n_{i+1} \Big) + \nonumber\\
&+&  z^2 (1 - \zeta_{N-1})  \sum_{\{n_i\}} 
z^{\sum_{i = 1}^{N-2}  n_i} \; \prod_{i = 1}^{N - 3} \Big(1 - \zeta_i \; n_i \; n_{i+1} \Big)  
\Big(1 - \zeta_{N - 2} \; n_{N-2} \Big) = \nonumber\\
&=& z Z_{N-2}(\zeta) + z (1 - \zeta_{N-1}) Z_{N-1}'(\zeta)
\end{eqnarray}
Now, recursion in Eq.(\ref{1}) 
allows us to eliminate $Z_N'(\zeta)$ in Eq.(\ref{2}).
 From Eq.(\ref{1}) we have 
$Z_N'(\zeta) = Z_N(\zeta) - Z_{N-1}(\zeta)$, and consequently, we find from
Eq.(\ref{2}) that the unconstrained 
partition function  $Z_N(\zeta)$ in Eq.(\ref{partition}) obeys the following recursion
\begin{equation}
\label{rec1}
Z_N(\zeta) = \Big(1 + z (1 - \zeta_{N - 1})\Big) Z_{N-1}(\zeta) +
 z \zeta_{N - 1}  Z_{N-2}(\zeta),
\end{equation}
which is to be solved subject to evident
 initial conditions
\begin{equation}
Z_0(\zeta) \equiv 1 \;\;\; \text{and} \;\;\; Z_1(\zeta) \equiv 1 + z.
\end{equation}

A conventional way (see, e.g. Ref.\cite{5,55,56})
to study linear random three-term recursions is to reduce them 
to random maps by introducing the Ricatti variable of the form 
\begin{equation}
R_{N}(\zeta) = \frac{Z_N(\zeta)}{Z_{N-1}(\zeta)}
\end{equation}
In terms of this variable Eq.(\ref{rec1}) becomes
\begin{equation}
\label{rec2}
R_N(\zeta) = \Big(1 + z (1 - \zeta_{N - 1})\Big)  +
 \frac{z \zeta_{N - 1}}{R_{N-1}(\zeta)}, \;\;\; \text{with} \;\;\;  R_1(\zeta) \equiv R_1 = 1 + z,
\end{equation}
which represents a random homographic relation.
Once $R_N(\zeta)$ is defined for arbitrary $N$,
the partition function $Z_N(\zeta)$ can be readily determined as the product,
\begin{equation}
Z_N(\zeta) = \prod_{i = 1}^N  R_i(\zeta),
\end{equation}
and hence, the disorder-average logarithm of the partition function 
will be obtained as
\begin{equation}
\label{sum}
\Big< \ln Z_N(\zeta)\Big>_{\zeta} = \sum_{i = 1}^N \Big< \ln R_i(\zeta) \Big>_{\zeta}
\end{equation}

Before we proceed further on, we note 
that recursion schemes of quite a similar form 
have been discussed already in the 
literature in different contexts. 
In particular, two decades ago 
Derrida and Hilhorst \cite{5} (see also Ref.\cite{99} 
for a more general discussion) 
have shown that such recursions occur
in the analysis of 
the Lyapunov exponent $F(\epsilon)$
of the product of random $2 \times 2$ matrices of the form
\begin{equation}
\label{pp}
F(\epsilon) = \lim_{N \to \infty} \frac{1}{N}\Big< \ln\Big({\rm Tr}\Big[\prod_{i=1}^{N} 
\begin{pmatrix}
1 & \epsilon \\
z_i \epsilon & z_i 
\end{pmatrix}
\Big]\Big)\Big>_{\{z_i\}},
\end{equation}
where $z_i$ are independent positive random numbers with a given probability distribution $\rho(z)$.  
Equation (\ref{pp}) is related, for instance, to the disorder-average 
free energy of  
an Ising chain with nearest-neighbor interactions in a random
magnetic field, and appears in the solution of a 
two-dimensional Ising model with row-wise random 
vertical interactions \cite{6}, the role of
$\epsilon$ being played by the wavenumber $\theta$. 
The recurence scheme in Eq.(\ref{rec2}) 
emerges also in such 
an interesting context as the problem 
of enumeration of primitive words with 
random errors in the locally free 
and braid groups \cite{100}.   
Some other examples of physical systems 
in which the recursion in Eq.(\ref{rec2}) 
appears 
can be found in Ref.\cite{55}. 

Further on, Derrida and Hilhorst \cite{5} 
have demonstrated 
that $F(\epsilon)$ can be expressed as
\begin{equation}
F(\epsilon) = \lim_{N \to \infty} \frac{1}{N} \sum_{i=1}^{N} \Big<\ln R'_i\Big>_{\{z_i\}},
\end{equation}
where  $ R'_i$ are defined through the recursion
\begin{equation}
\label{rec3}
 R'_i = 1 +z_{i-1} + z_{i-1}(\epsilon^2 - 1)/R'_{i-1}, \;\;\; 
\text{with} \;\;\; R'_1 = 1.
\end{equation} 
Moreover,  
they have shown that the model admits an exact solution 
 when 
\begin{equation}
\label{dist}
\rho(z) = (1 - p) \delta(z) + p \delta(z - y),
\end{equation}
i.e. when similarly to the model under study,
 $z_i$ are independent, random two-state 
variables assuming only two values - 
$y$ with probability $p$ and $0$ with probability $1 - p$. 
Supposing that when $i$ increases, a stationary probability
distribution $P(R')$ of the $R'_i$ independent of $i$ exists \cite{8},
Derrida and Hilhorst \cite{5} have found the following exact result:
\begin{eqnarray}
\label{k}
F(\epsilon) &=& p \ln(1 + b) - p (2 - p) \ln(1 + b \frac{y - b}{1 - b y}) + \nonumber\\ 
&+& (1 - p)^2 \sum_{N = 1}^{\infty} p^N \ln(1 + b \Big(\frac{y - b}{1 - b y}\Big)^{N + 1}),
\end{eqnarray}
where
\begin{equation}
\label{k1}
b = 1 + \frac{(1 - y)^2}{2 \epsilon^2 y} 
\Big[1 - \Big(1 + 4 \frac{\epsilon^2 y}{(1 - y)^2}\Big)^{1/2}\Big].
\end{equation}

\section{Disorder-average pressure.}

We turn now back to our recursion scheme in Eq.(\ref{rec2}) and notice that setting
\begin{equation}
R_i(\zeta) = (1 + z) \; R'_i,
\end{equation}
and choosing
\begin{equation}
\label{def}
y = - \frac{z}{1 + z} = - n^{(lan)}, \;\;\; \text{and} \;\;\; \epsilon^2 = \frac{z}{1 + z} = n^{(lan)},
\end{equation}
makes the recursion schemes in Eqs.(\ref{rec2}) and (\ref{rec3}) identic! 
Consequently, the disorder-average pressure
per site in our random catalytically-activated  reaction/adsorption model can be expressed as
\begin{equation}
\label{press}
P^{(quen)} \equiv \frac{1}{\beta} \ln(1 + z) + \frac{1}{\beta} F(\epsilon),
\end{equation}
where $F(\epsilon)$ is the 
Lyapunov exponent of the product of random $2 \times 2$ matrices in Eq.(\ref{pp}), in which
$\epsilon$ and $z_i$ are defined by Eqs.(\ref{dist}) and (\ref{def}). 

Note next that the first 
term on the right-hand-side
of Eq.(\ref{press}) is a trivial Langmuir result for the $p = 0$ case 
(adsorption without reaction) which would entail $n^{(quen)} = z/(1 + z)$, Eq.(\ref{densi}).
Hence, all non-trivial, disorder-induced behavior is
 embodied in the Lyapunov exponent $F(\epsilon)$.

The disorder-averaged pressure per site for the random reaction/adsorption 
model under study 
can be thus readily obtained
from Eqs.(\ref{k}) and (\ref{k1}) by 
defining 
the parameters 
$y$ and $\epsilon$ as prescribed in Eq.(\ref{def}).
This yields the following exact result:
\begin{eqnarray}
\label{array5}
\beta P^{(quen)} &=& \ln(\phi_z) - (1 - p) \ln\Big(1 - \omega^2\Big) + \nonumber\\
&+& \frac{(1 - p)^2}{p} \sum_{N = 1}^{\infty} p^{N } 
\ln\Big(1 - (-1)^N  \omega^{N + 2}\Big), 
\end{eqnarray}
where
\begin{equation}
\phi_z = \frac{1 + \sqrt{1 + 4 z}}{2},
\end{equation}
and
\begin{equation}
\label{omega}
\omega = \frac{\sqrt{1 + 4 z} - 1}{\sqrt{1 + 4 z} + 1} = z/\phi_z^2 = 1 - \frac{1}{\phi_z}
\end{equation}
Note that $\phi_z$ obeys $\phi_z (\phi_z - 1) = z$; hence,  
$\phi_{z=1}= (\sqrt{5} + 1)/2$ is just the "golden mean".

\section{Asymptotic behavior of the disorder-average pressure, mean density and the compressibility.}

Consider first the asymptotic behavior of 
$P^{(quen)}$ in the small-$z$ limit. To do this, it is expedient to use
another representation of $P^{(quen)}$. After some straightforward calculations, 
one can cast $P^{(quen)}$ in Eq.(\ref{array5}) into the form:
\begin{equation}
\beta P^{(quen)} = \frac{(1 - p)}{p} \sum_{n = 0}^{\infty} p^n {\cal F}_n,
\end{equation}
where ${\cal F}_n$ denote
 natural logarithms of 
the Stieltjes-type continued fractions of the form
\begin{eqnarray}
\label{F}
{\cal F}_n = \ln\left(1 + 
\frac{z}{{\displaystyle 1+\frac{z}{{\displaystyle
1+\frac{z}{{\displaystyle 1+\frac{\cdots}{{\displaystyle
1+z}}}}}}}}
\right).
\end{eqnarray}
Note now that in the limit $n \to \infty$, one has
\begin{equation}
\lim_{n \to \infty} {\cal F}_n = \ln(\phi_z) = \ln\Big(\frac{1 + \sqrt{1 + 4 z}}{2}\Big),
\end{equation}
i.e. ${\cal F}_n$ is the $n$-th approximant 
of $\ln(\phi_z)$; hence,  $P^{(quen)}$ can be thought of as 
the generating function of such approximants. Now,
one finds that for $z < 1$  the sequence of approximants 
converges quickly to $\ln(\phi_z)$; 
expanding 
the $n$-th approximant ${\cal F}_n$ into the Taylor series in powers of $z$, 
one
has that the 
first $n$ terms of such an 
expansion coincide with 
the first $n$ terms
of the expansion of $\ln(\phi_z)$, i.e.
\begin{equation}
\label{nnnn}
\ln(\phi_z) =  \ln\Big(\frac{1 + \sqrt{1 + 4 z}}{2}\Big) = - 
\frac{1}{2 \sqrt{\pi}} \sum_{n=1}^{\infty} \frac{(-1)^n \Gamma(n + 1/2)}{\Gamma(n + 1)} \frac{(4 z)^n}{n},
\end{equation}
Consequently,  ${\cal F}_n$ and ${\cal F}_{n-1}$ differ
 only by terms of order $z^n$, which signifies that convergence is 
good.
On the other hand,  
for $z \geq 1$ convergence becomes poor 
and one has to seek for a more suitable 
representation. As a matter of fact,  
already for $z = 1$ one has 
that in  
the limit $n \to \infty$ 
the approximant ${\cal F}_n$ tends to  $\ln(\phi_1)$, 
i.e. 
the logarithm of the 
"golden mean", which is known as the 
irrational number worst approximated by rationals. 

In the small-$z$ limit, we find then using an expansion in Eq.(\ref{nnnn})
that  $P^{(quen)}$ follows
\begin{eqnarray}
\label{pppp}
\beta P^{(quen)} = z - \Big(\frac{1}{2} + p\Big) z^2 + 
\Big(\frac{1}{3} + 2 p + p^2\Big) z^3 -
 \Big(\frac{1}{4} + \frac{7}{2} p + 4 p^2 + p^3 \Big) z^4 + {\cal O}(z^5).
\end{eqnarray}
Consequently, in the small-$z$ limit the mean density obeys:
\begin{equation}
\label{nn}
n^{(quen)} = z - (1 + 2 p) z^2 + \Big(1 + 6 p  + 3 p^2\Big) z^3 - 
\Big(1 + 14 p  + 16 p^2 + 4 p^2\Big)z^4  + {\cal O}(z^5),
\end{equation}
while the compressibility $k_T^{(quen)}$ is given by
\begin{equation}
\label{kk}
 \beta^{-1} k_T^{(quen)} =  
\frac{1}{z} + p (2 - p) z - 4 p (2 - p) z^2 +  3 p \Big( 8 - p - 2 p^2\Big) z^3 + {\cal O}(z^4).
\end{equation}
Note  that , the coefficients 
in the small-$z$ expansion coincide with the coefficients in the 
expansions of $P^{(lan)}$ and  $P^{(reg)}$ when we set in Eq.(\ref{pppp}) $p = 0$ or $p = 1$.

Now, we turn to the analysis of the large-$z$ behavior 
which
is a bit more complex than $z \ll 1$ case 
and requires understanding of the
asymptotic behavior of the sum
\begin{equation}
S = \sum_{N=1}^{\infty} p^N \ln\Big(1 - (-1)^N \omega^{N + 2}\Big)
\end{equation}
entering Eq.(\ref{array5}). We note first that in this sum 
the behavior of the terms with odd and even $N$ is quite different and we have to consider it separately.
Let
\begin{equation}
S_{odd} = \frac{1}{p} \sum_{N=1}^{\infty} p^{2 N} \ln\Big(1 + \omega^{2 N + 1}\Big)
\end{equation}
denote the contribution of the terms with odd $N$. Note that when $z \to \infty$ (i.e. $\omega \to 1$) 
the sum $S_{odd}$ tends to $p \ln(2)/(1 - p^2)$. The corrections to this limiting behavior can be defined as follows. Expanding
$\ln\Big(1 + \omega^{2 N + 1}\Big)$ into the Taylor series in powers of $\omega$ and 
then, using the definition $\omega = 1 - 1/\phi_z$ and the binomial expansion, 
we construct a series in the inverse powers of $\phi_z$:
\begin{eqnarray}
\label{klm}
S_{odd} &=& \frac{p}{1 - p^2} \ln(2) - \frac{1}{2} \frac{p (3 - p^2)}{(1 - p^2)^2} \frac{1}{\phi_z} + \nonumber\\
&+& \frac{1}{8} \frac{p (3 + 6 p^2 - p^4)}{(1 - p^2)^3} \frac{1}{\phi_z^2} + 
 \frac{1}{24} \frac{p (15 + 10 p^2 - p^4)}{(1 - p^2)^3} \frac{1}{\phi_z^3} + {\cal O}\Big(\frac{1}{\phi_z^4}\Big) 
\end{eqnarray}
Note that this expansion is only meaningful when $\phi_z \gg (1 - p)^{-1}$, ($z \gg (1 - p)^{-2}$), which signifies that $p = 1$ is a special
point. 

Further on, plugging into the latter expansion the 
definition of $\phi_z$, $\phi_z = (1 + \sqrt{1 + 4 z})/2$, 
we obtain the following expansion in the inverse powers of the activity $z$:
\begin{eqnarray}
\label{mmm}
S_{odd} &=& \frac{p}{1 - p^2} \ln(2) - \frac{p}{2} \frac{(3 - p^2)}{(1 - p^2)^2} \frac{1}{z^{1/2}} + \nonumber\\
&+& \frac{p}{8} \frac{(9 - 2 p^2 + p^4)}{(1 - p^2)^3} \frac{1}{z} + 
 \frac{p}{48} \frac{(3 - 4 p^2 + p^4)}{(1 - p^2)^3} \frac{1}{z^{3/2}} + {\cal O}\Big(\frac{1}{z^2}\Big) 
\end{eqnarray}

Consider next the sum
\begin{equation}
\label{even}
S_{even} =  \sum_{N=1}^{\infty} p^{2 N} \ln\Big(1 - \omega^{2 N + 2}\Big),
\end{equation}
which represents the contribution of terms with even $N$. Note 
that in contrast to the behavior of $S_{odd}$, the sum in Eq.(\ref{even})
diverges when $z \to \infty$ ($\omega \to 1$). 
Since $1 - \omega^{2 N + 2} \sim 1 - \omega$ when $\omega \to 1$, we have
that in this limit the leadin behavior of $S_{even}$ is described by
\begin{equation}
\label{even2}
S_{even} \sim \frac{p^2}{1 - p^2} \ln(1 - \omega).
\end{equation} 
To obtain several correction terms we make use of one of Gessel's expansions
\cite{gessel}:
\begin{equation}
\label{gessel}
\ln\Big(\frac{2 (N + 1) x}{1 - (1 - x)^{2 N + 2}}\Big) = \sum_{k = 1}^{\infty} g_k(2 N + 2) \frac{(-1)^k x^k}{k},
\end{equation}
where $g_k(2 N + 2)$ are the Dedekind-type sums of the form
\begin{equation}
g_k(2 N + 2) = \sum_{\zeta^{2 N + 2} = 1, \zeta \neq 1} \frac{1}{\Big(\zeta - 1\Big)^{k}},
\end{equation}  
where the summation extends over all $\zeta$ being the $(2 N + 2)$-th roots of unity (with $\zeta = 1$ excluded).
As shown in Ref.\cite{gessel}, the weights $g_k(2 N + 2)$ are 
polynomials in $N$ of degree at most $k$ with rational coefficients.
Next, setting $x = 1/\phi_z$ in the expansion in Eq.(\ref{gessel}), plugging it to Eq.(\ref{even}) 
and performing summations over $N$, 
we find that $S_{even}$ can be written down as
\begin{equation}
S_{even} = - \frac{p^2}{1 - p^2} \ln(\phi_z) + \frac{p^2}{1 - p^2} \ln(2) + s_p - \sum_{k=1}^{\infty} G_k(p) \frac{(-1)^k}{k \phi_z^k},
\end{equation}
where $s_p$ is an infinite series of the form \footnote{Note that $s_p$ shows a non-analytic behavior when $p \to 1$. This function can be represented as
\begin{equation}
s_p = - \frac{1}{1 - p^2} \ln(1 - p^2) - \frac{p^2}{1 - p^2} \sum_{n = 2}^{\infty} \frac{(-1)^n}{n} \Phi(p^2,n,1),
\end{equation}
where $\Phi(p^2,n,1)$ are the Lerch transcedents, $\Phi(p^2,n,1) = \sum_{l = 0}^{\infty} (1 + l)^{-n} p^{2 l}$. It is 
straightforward to find then that $s_p =  - \frac{1}{1 - p^2} \ln(1 - p^2) - \frac{\gamma}{1 - p^2} + {\cal O}(\ln(p))$,
where $\gamma$ is the Euler constant.}
\begin{equation}
s_p = \sum_{N = 1}^{\infty} p^{2 N} \ln(N + 1), 
\end{equation}
while $G_k(p)$ are the generating functions of the polynomials $g_k(2 N + 2)$:
\begin{equation}
G_k(p) = \sum_{N=1}^{\infty} g_k(2 N + 2) p^{2 N}.
\end{equation}
Inserting next the definition of $\phi_z$, we find the following explicit asymptotic expansion
\begin{eqnarray}
\label{j2}
S_{even} &=& - \frac{1}{2} \frac{p^2}{1 - p^2}  \ln(z) + \frac{p^2}{1 - p^2} \ln(2) + s_p + \nonumber\\
&-& \frac{p^2 (2 - p^2)}{(1 - p^2)^2} \frac{1}{z^{1/2}} + \frac{p^2 (21 - 18 p^2 + 5 p^4)}{24 (1 - p^2)^3} \frac{1}{z} +
\frac{p^2 (2 - p^2)}{24 (1 - p^2)^2} \frac{1}{z^{3/2}} + {\cal O}\Big(\frac{1}{z^{2}}\Big)
\end{eqnarray}
Finally, 
combining the expansions in Eqs.(\ref{array5}), (\ref{mmm}) and (\ref{j2}), we find the following large-$z$ expansion 
for the disorder-averaged pressure $P^{(quen)}$:
\begin{eqnarray}
\label{presq}
\beta P^{(quen)} &=& \frac{1}{1 + p} \ln(z) - \frac{(1 - p)^2}{(1 + p)} \ln(2) + \nonumber\\ 
&+& \frac{(1 - p)^2}{p} s_p + 
\frac{1}{6} \frac{6 + 3 p - p^3}{(1 + p)^2 (1 - p^2)} \frac{1}{z} + {\cal O}\Big(\frac{1}{z^2}\Big),
\end{eqnarray}
which yields
\begin{equation}
\label{densit}
n^{(quen)} = \frac{1}{1 + p} - \frac{1}{6} \frac{6 + 3 p - p^3}{(1 + p)^2 (1 - p^2)} \frac{1}{z} + {\cal O}\Big(\frac{1}{z^2}\Big),
\end{equation}
and
\begin{equation}
\label{cc}
\beta^{-1} k_T^{(quen)} = 
\frac{1}{6} \frac{6 + 3 p - p^3}{(1 + p) (1 - p^2)} \frac{1}{z} +
\frac{1}{36} \frac{p \Big(6 + 3 p - p^3\Big)^2}{(1 + p)^2 (1 - p^2)^2} \frac{1}{z^2} + {\cal O}\Big(\frac{1}{z^3}\Big).
\end{equation}
Note that asymptotic expansions in Eqs.(\ref{presq}), (\ref{densit}) and (\ref{cc}) are only meaningful
for $z \gg (1 - p)^{-2}$ and thus exclude the
completely catalytic $p = 1$ case. At this special point $p \equiv 1$ we find from Eqs.(\ref{ord})
that the pressure and mean density exhibit a non-analytic dependence on $1/z$:
\begin{equation}
\label{4i}
\beta P^{(reg)} = \frac{1}{2} \ln(z) + \frac{1}{2 z^{1/2}} - \frac{1}{48
z^{3/2}} + \frac{3}{1280 z^{5/2}} + {\cal O}\Big(\frac{1}{z^{7/2}}\Big),
\end{equation}
and
\begin{eqnarray}
n^{(reg)} = \frac{1}{2} - \frac{1}{4 z^{1/2}} + \frac{1}{32 z^{3/2}} -
\frac{3}{512 z^{5/2}} + {\cal O}\Big(\frac{1}{z^{7/2}}\Big),  
\end{eqnarray} 
which differs substantially from the asymptotical
 behavior in the $p < 1$ case, Eqs.(\ref{presq}) and 
(\ref{densit}). This happens apparently 
because the bulk contribution to the
disorder-average pressure in  Eqs.(\ref{presq}) comes 
from the intervals devoid of 
the catalytic segments, in which reactions can not takes place and the 
mean density $n \sim 1$ in accordance with the Langmuir
adsorption/desorption mechanism. 
Such intervals exist for any  
$p$ strictly less than unity; their contribution 
 vanishes only when $p \equiv 1$.

\section{Conclusions.}

To conclude, in this paper we have presented an exact solution of a 
random catalytic reaction/adsorption model, 
appropriate to the 
 situations 
with the catalytically-activated reactions
on polymer chains containing randomly placed catalytist.
More specifically, we have considered
here the $A + A \to 0$ reaction 
on a one-dimensional regular lattice which is brought 
in contact with a reservoir of $A$ partilces.
Some portion of the intersite intervals  on the regular lattice 
was supposed to possess special "catalytic" properties such that
they induce an immediate
 reaction $A + A \to 0$, as soon as
two $A$ particles land onto two vacant sites at the
extremities of the  catalytic segment, or an 
$A$ particle lands onto a vacant site
while the site at the 
other extremity of the
catalytic segment is already occupied 
by another $A$ particle. 
For $quenched$   random distribution of 
the catalytic segments, 
we have determined exactly the disorder-averaged pressure per site
and have shown 
that it can be represented as  a 
sum of a
Langmuir-type contribution and a reaction-induced term. The latter 
can be expressed 
as the Lyapunov exponent
of a product of random two-by-two matrices, obtained by Derrida and Hilhorst
\cite{5}. Explicit asymptotic expansions for the mean particle
density and the compressibility were also derived.

\end{document}